\documentclass[letterpaper]{ieeeconf}
\usepackage{amsmath}
\usepackage{graphicx,subfigure,wrapfig}
\usepackage{algorithm,algpseudocode,float}
\usepackage{xspace}
\usepackage[table]{xcolor}

\title{Computational Social Dynamics: Analyzing the Face-level Interactions in a Group\date{}}
\author{\parbox{16cm}{\centering
		{\large Nicholas Watkins and Ifeoma Nwogu}\\
		{\normalsize
			Department of Computer Science, Rochester Institute of Technology\\
			Rochester, NY}}}
\graphicspath{{images/}}
\newcommand{\ie}{\emph{i.e.}\xspace}
\newcommand{\eg}{\emph{e.g.}\xspace}
\newcommand{\etc}{\emph{etc.}\xspace}
\newcommand{\etal}{\emph{et~al.}\xspace}

\begin{document}

\maketitle

\begin{abstract}
	Interactional synchrony refers to how the speech or behavior of two or more people involved in a conversation become more finely synchronized with each other, and they can appear to behave almost in direct response to one another. Studies have shown that interactional  synchrony is a hallmark of relationships, and is produced as a result of rapport. 
	In this work, we use computer vision based methods to extract nonverbal cues, specifically from the face, and develop a model to measure interactional synchrony based on those cues. This paper illustrates a novel method of constructing a dynamic deep neural architecture, specifically made up of intermediary long short-term memory networks (LSTMs), useful for learning and predicting the extent of synchrony between two or more processes, by emulating the nonlinear dependencies between them. On a synthetic dataset, where pairs of sequences were generated from a Gaussian process with known covariates, the architecture could successfully determine the covariance values of the generating process within an error of 0.5\% when tested on 100 pairs of interacting signals. On a real-life dataset involving groups of three people, the model successfully estimated the extent of synchrony of each group on a scale of 1 to 5, with an overall prediction mean of $2.96\%$ error when performing 5-fold validation, as compared to 26.1\% on the random permutations serving as the control baseline.   
\end{abstract}

\section{Introduction}
Research in Social Psychology has shown extensively that in cohesive groups, individuals typically mirror each other's facial expressions and facial articulations \cite{mirror}. Furthermore, in groups with established rapport, typically individuals provide feedback to others' opinions using facial features. Thus the ability to measure the extent of interactional synchrony among the members in a group can be used as a metric for cohesion or rapport within that group. 

Developing and maintaining rapport is a critical component of successful interactions in different social settings. For example, it has been shown in that in organizational settings, using techniques of rapport management in the interpersonal communication between leaders and team members can lead to building more positive relationships in the organization, which in turn can result in higher quality of work\cite{AlabamaStudy}. In a separate study from Japan\cite{prisoners_2017} that investigated prisoners' reasons for confessions when being interrogated, participants who experienced a relationship-focused interviewing style, which stressed active listening and rapport-building while talking about their criminal incidents directly, were more likely to confess. In yet another study on interracial mentoring relationships, findings suggested that affect and rapport were the key features in facilitating more positive outcomes\cite{leitner_2018}. In general, rapport management is a key component of social dynamics\footnote{Social dynamics refers to the behavior of groups, which results from the interactions of individual group members. It also refers to the study of the relationship between individual-level interactions and group-level behaviors.}
and a quantitative measure of the extent of rapport can prove to be very useful in difference social constellations.

We expect pairs or groups of individuals with high cohesion and established rapport to exhibit the \emph{mirroring effect}, a form of interactional synchrony. The mirroring effect is a phenomenon which occurs when individuals mimic each other's behavior subconsciously to gain and keep rapport\cite{mirror}. Thus, we expect the cohesion of a group to directly correspond the extent of synchrony occurring. Interactional synchrony can be manifested in various forms other than mirroring which is the most common. These include entrainment at the voice prosody level, convergence (when people in a group initially behave very differently from each other but eventually begin to behave similarly), mimicking, which is a type of mirroring but with time delays, \etc. 

To this end, to better understand interactional synchrony, we study coupled or interacting processes, where given two or more time series signals, our goal is to determine if and how much they are interacting. This general problem has numerous applications such as modeling human impact in social dynamics (as described above), explaining weather patterns, studying financial markets and their dependences, \etc In this work, we focus on studying how well individuals are interacting with each other in a given social context. 

We abstract our concept of processes by examining measurements over time (\eg once every second) and refer to these sets of time series $X$ and $Y$. We say two time series $X, Y$ are synchronous if their corresponding processes are interacting, \ie if the patterns in $X$ are influenced by $Y$ or vice versa. We assume that interaction always has an impact on the process.

\section{Past Related Work on Computational Synchrony}
While the notion of interactional synchrony has been studied extensively in the Social Psychology literature, much less work has been reported in the computational analysis literature on this subject. Delaherche \etal \cite{syncSurvey12} presented an extensive survey of synchrony evaluation from a multidisciplinary perspective, focusing on psychologists' coding methods, non-computational evaluation and early machine learning techniques. 

Rudimentary forms of synchrony analysis has long been performed using covariance measures in statistics, however, this only works if the processes have a liner relation.  But in real-life settings, such interactions can be nonlinear. Although most covariate time series are synchronous, not all synchronous data is covariate, such as is the case when synchronous data are nonlinearly coupled. Additionally, synchronous data can have a similar noise pattern (with an underlying independent trend) or have a nonuniform delay. These cases would all fail under traditional variance analysis. Synchronicity  analysis has also been previously approached by using coupled hidden Markov models (cHMMs) \cite{brand_oliver_pentland, cHMM2} to classify taichi movements, under the assumption that different parts of the body moving in taichi will be synchronous to each other. Although cHMMs are powerful, they cannot model irregular delays between the two sequences. Furthermore, due to the stringent Markovian assumption\footnote{The $k^{th}$ order Markov assumption states that only the $t-k, t-k+1, ..., t-1, t$ measurements impact the $t+1$ measurement}, 
they are often heavily augmented to be useful. Li \etal \cite{LiCH15} presented a supervised model used to predict the outcomes of video-conferencing conversations in the context of new recruit negotiations, again making the Markov assumption on interactional synchrony between the interlocutors. Yu \etal \cite{Yu2013AutomatedAO} presented a technique to investigate interactive synchrony in facial expressions and showed using the Pearson's correlation measure, that synchrony features were effective at detecting deception; Hammal \etal \cite{Hammal13} evaluated the temporal coordination of head movements in couples with history of interpersonal violence and Chu \etal  \cite{chu2015unsupervised} developed a search-based technique for unsupervised, accelerated, multi-synchrony detection. In much of the model based works described, first-order Markov assumptions are made for modeling interactional synchrony.

Group cohesion analysis (not in the context of synchrony) has been previously approached by using using SVMs on extracted audiovisual features from group meeting videos labeled as having high or low cohesion by human annotators  \cite{hung_gatica-perez_2010}. In this paper we describe our work on studying group cohesion using a neural network model, specifically the long short-term memory (LSTM). Our method is advantageous over the aforementioned ones in that our model uses less preset features, we are injecting less bias into the model and it is well-suited to classify, process and predict time series data given time lags of unknown size and duration between events.

\section{Method}
In this section, we present our techniques for synthetic data generation, feature extraction and selection from the real-life dataset and interactional synchrony measurement via the LSTM. We also describe the process of obtaining a form of ground truth via human annotations, since labels were not provided for the data. 
LSTMs are a class of recursive neural networks (RNN) which use memory cells to ``remember'' prior computations. LSTMs and RNNs are frequently used for time series prediction and classification. The memory cells of LSTM networks cause the time series they generate to adhere to higher orders of the Markov property. They are therefore capable of learning and modeling arbitrary sequential functions (given the neural topology is sufficiently complex).

\begin{figure}[h]
\centering
		\vspace*{-10pt}
		\includegraphics[width=0.9\linewidth]{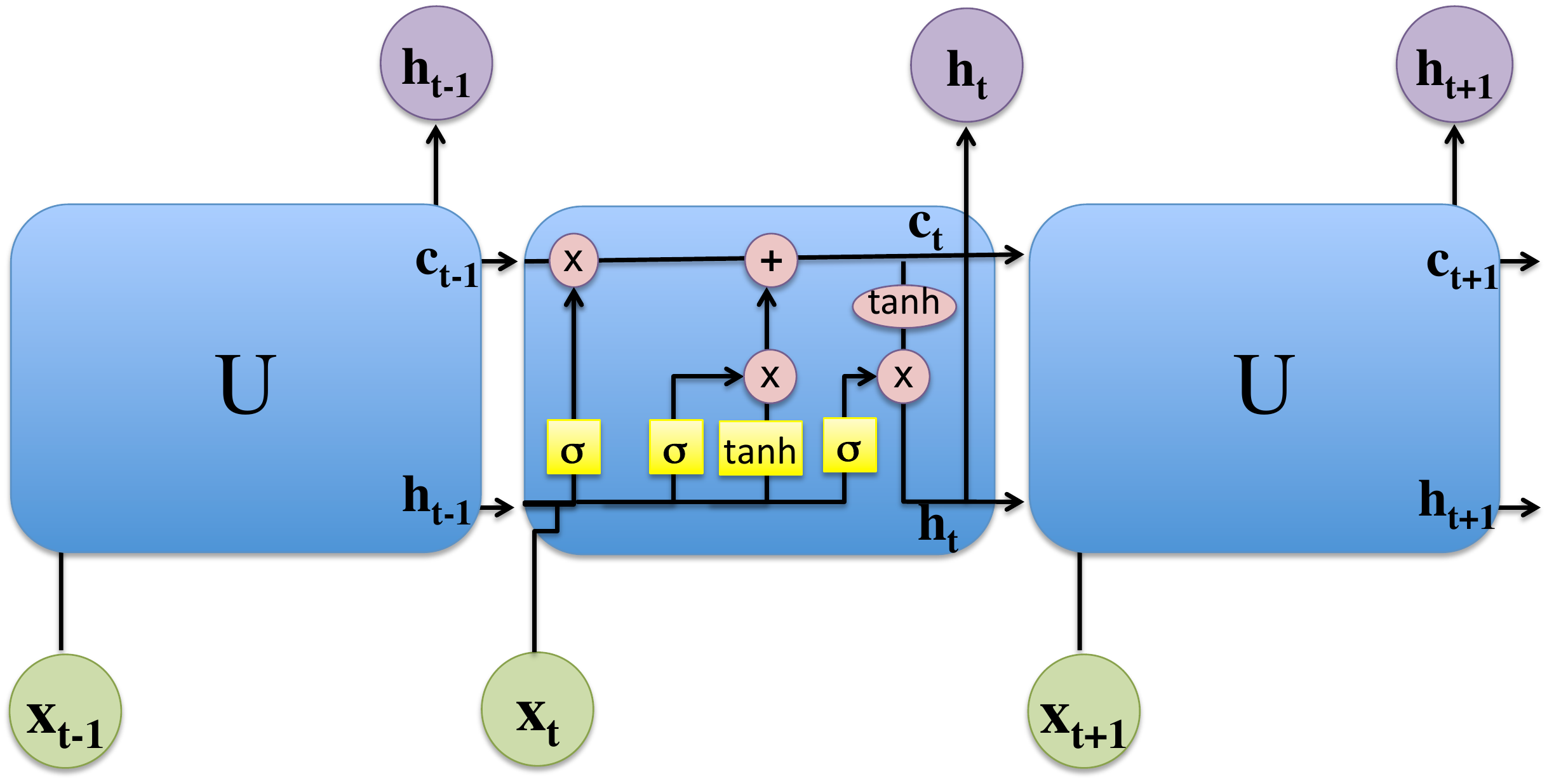}
		\caption{An unrolled LSTM network showing the repeating module $U$ with four interacting layers internally, along with inputs $\{x_i\}$ and outputs $\{h_i\}$.}\label{fig:lstm_basic}
\end{figure}

\subsection{Synthetic data}
To sufficiently ensure our approach would be successful on a variety of different applications, we generated synthetic data with desired levels of synchronousity. We took inspiration from Jamali's method of generating correlated Gaussian processes \cite{datagen} and added a post processing step to ensure the data need not necessarily be correlated (but merely synchronous). We also implemented a rudimentary tread feature that allows us to prescribe a trend to both time series. 
The extended version of the algorithm is described in more detail in Algorithm \ref{alg:datagen}. 
\begin{algorithm}[H]
	\caption{Data generating process}\label{alg:datagen}
	\begin{algorithmic}[1]
		\Procedure{DataGen}{$len, C, f_1(t,x), f_2(t,x), delay$}
		\State $S_{xx} = fft(C_{xx})$
		\State $S_{yy} = fft(C_{yy})$
		\State $S_{xy} = fft(C_{xy})$
		\State $u      = SampleGaussian(len)$
		\State $v      = SampleGaussian(len)$
		\State $U      = fft(u)$
		\State $V      = fft(v)$
		\State $\beta   = 1$
		\State $\alpha  = \overline{\beta} + \cos^{-1}\left(\frac{S_{xy}}{\sqrt{Sxx * Syy}}\right)$
		\State $A_q    = \sqrt{S_{xx}} * \cos(\alpha)$
		\State $B_q    = \sqrt{S_{xx}} * \sin(\alpha)$
		\State $C_q    = \sqrt{S_{yy}} * \cos(\alpha)$
		\State $D_q    = \sqrt{S_{yy}} * \sin(\alpha)$
		\State $X      = A_q * U^T + B_q * V^T$
		\State $Y      = C_q * U^t + B_q * V^T$
		\State $x^*    = |ifft(X)|$
		\State $y^*    = |ifft(Y)|$
		\State $x      = x^*[1+delay: ]$
		\State $y      = y^*[1:-delay]$
		\State Return $[f_1(1,x_1),...,f_1(len,x_{len})],$ \\\hspace*{15.5mm}$[f_2(1,y_1),...,f_2(len,y_{len})]$
		\EndProcedure
	\end{algorithmic}
\end{algorithm}

\begin{figure}
	\centering
	\includegraphics[width=0.6\linewidth]{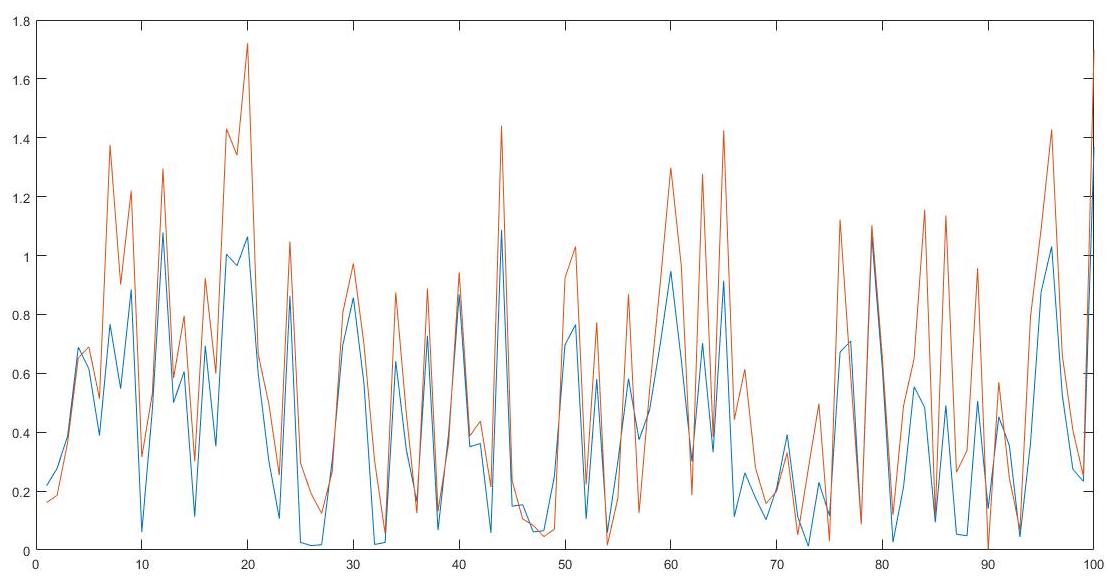} \hspace*{-4mm}\\
	\includegraphics[width=0.6\linewidth]{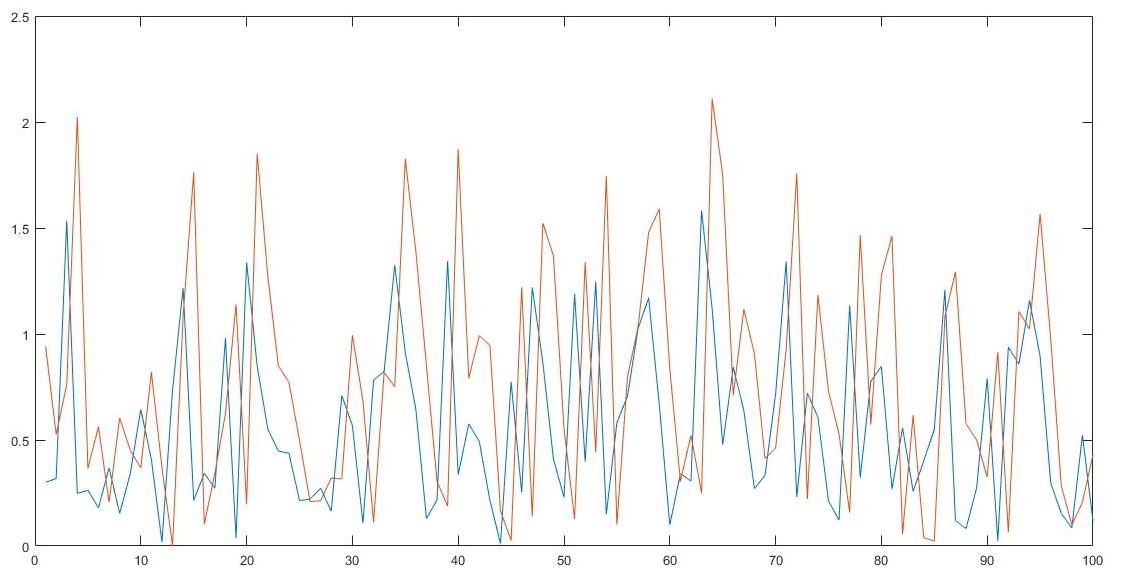} \hspace*{-4mm} \\
	\includegraphics[width=0.6\linewidth]{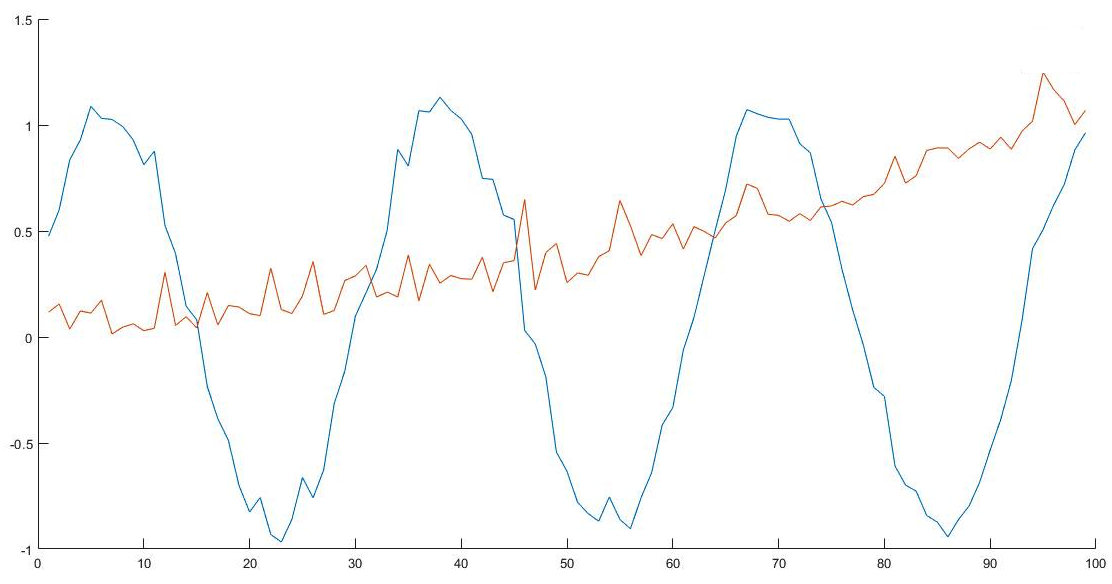} \\
	\caption{Examples of synchronized pairs of random variable sequences synthetically generated.}\label{fig:datagen}
\end{figure}

Examples of the outputs of the data generating process for a pair of synchronized 100-length sequences are shown in Figure \ref{fig:datagen}. The left image is a basic pair of stationary synchronized sequences: the middle image shows another pair of synchronized sequences, but shifted by one unit of time - note that the sequences are not exactly the same, rather they are random sequences generated by the same stochastic process.  The right image shows a pair of synchronized sequences where one of the signals is quasi-periodic and the other is trending
If the covariance between the two signals is given as: 
$
\begin{bmatrix} \Phi_{11} & \Phi_{12} \\ \Phi_{21} & \Phi_{22} \end{bmatrix}
$
and we train an ample number of signal pairs with the neural network structure presented in Section \ref{sec:model} using a regression loss on $\Phi_{12}$, we are able to effectively recover this value back when tested with newly generated signals with known covariances. Details of the experiment and results are given in Section \ref{sec:testing}.

\subsection{The Sayette Group Formation Task dataset}
\begin{figure}[h]
	\centering
	\begin{tabular}{cc}
		\includegraphics[width=0.47\linewidth]{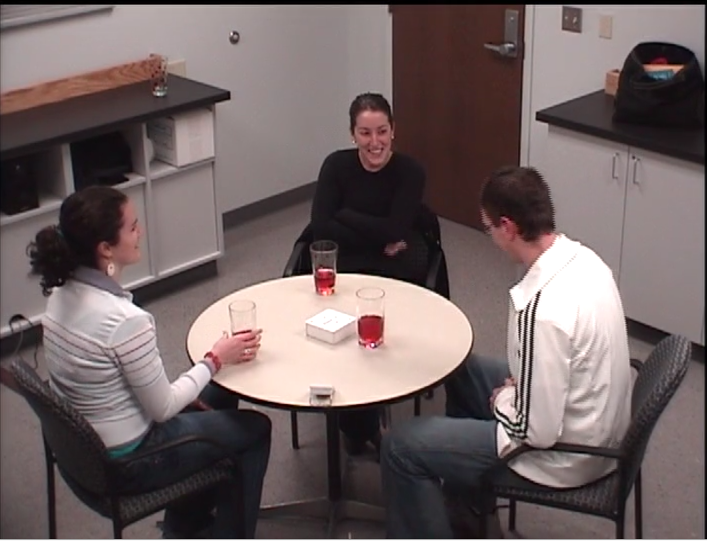} &
		\includegraphics[width=0.47\linewidth]{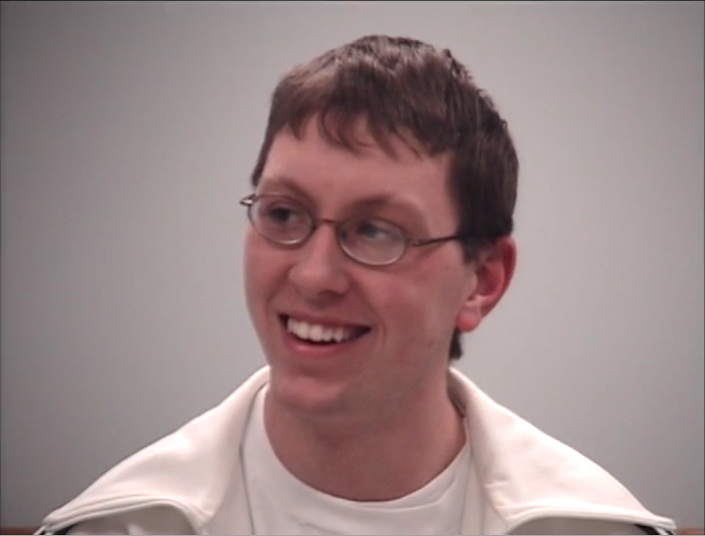} \\
		\includegraphics[width=0.47\linewidth]{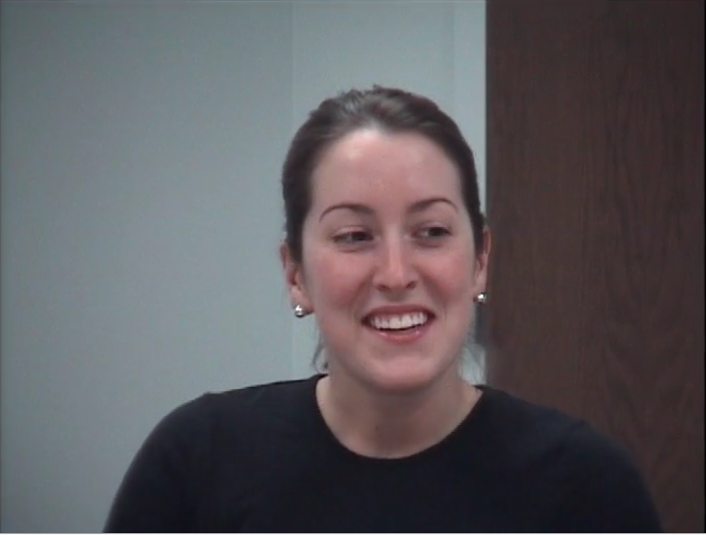} &
		\includegraphics[width=0.47\linewidth]{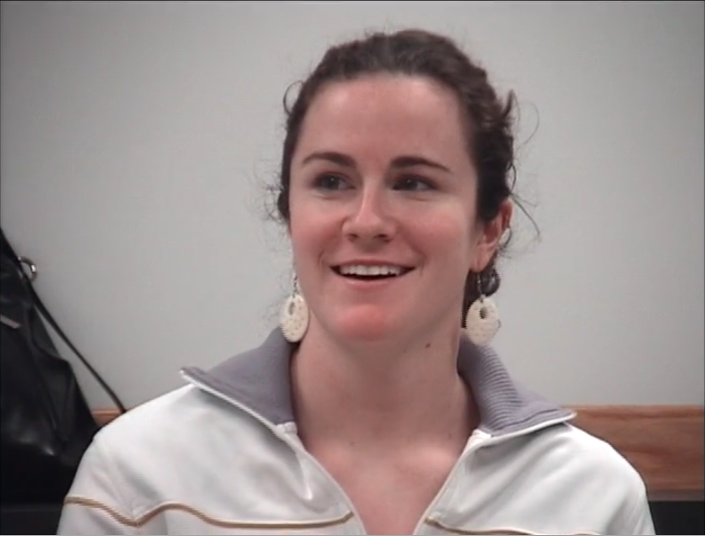} \\									
	\end{tabular}
	\caption{Examples of video frames from the dataset - group level and participant face level.}\label{fig:dataset}
\end{figure}

We are testing this approach on a dataset obtained from Girard \etal \cite{Girard17FG_GFT} which was drawn from a larger study on the impact of alcohol on group formation processes \cite{Sayette12}. All participants in the study were previously unacquainted and met for the first time at the experiment. They were instructed to consume a beverage and then engage with two other study participants. The groups of interlocutors were made up of three such subjects who were engaged for about 30-40 minutes of unstructured interactions. The data provided for this study focused on a 1 minute portion of the entire video where the collectors believed the participants in the group had become sufficiently acquainted with each other. Unfortunately, the associated labels for the videos were not provided. Separate wall-mounted cameras faced each participant and another camera captured the overall group interaction, resulting in a total of four videos - one at the overall group level showing body movements and three at the individual level showing mainly the face. Figure \ref{fig:dataset} shows sample frames from one of the videos depicting the overall group interaction as well as the faces of individual participants in the group. The database contains a total of 172,800 frames, with 1,800 frames for each of 96 participants.

\subsubsection{Data Processing}
\begin{figure}
	\centering
	\includegraphics[width=0.5\linewidth]{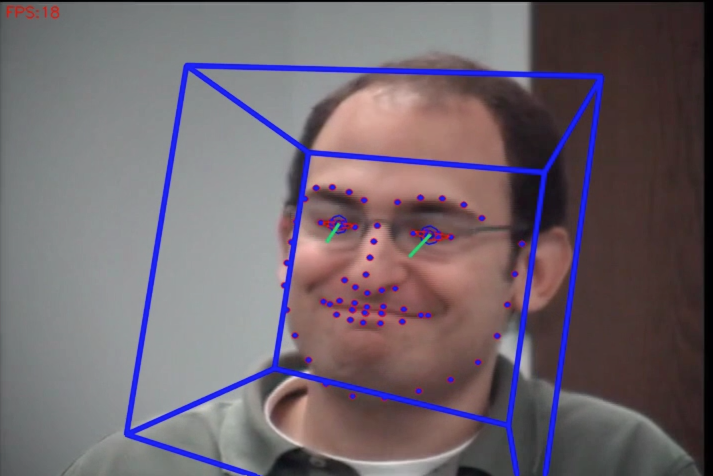}
	\caption{OpenFace output on a sample frame showing the facial landmarks used to compute action units, head pose and eye gaze.}
	\label{fig:openface}
\end{figure}
Each individual's facial expression was parsed using OpenFace \cite{openFace13}, an open-source toolkit capable of facial landmark detection, head pose estimation, facial action unit recognition, and eye-gaze estimation. Action units (AUs) correspond to various muscle groups on the face and can range from being fully activated to not activated\cite{FACS}. For each trio in a group, we took the top 3 most active action units. We defined
the activity of an action unit as the mean average deviation. The amount of mirroring directly corresponds to our definition of synchrony (as increased synchrony between the facial behavior in pairs of individuals indicates mirroring); thus, for our neural network to detect mirroring, it must detect synchrony (by proxy). 

We took these action units and measured them through time, thus obtaining a collection of behavioral signals (based on the face). These signals can be seen as a high order coupled Markov process as individual channels interacted with each other. 

\subsubsection{Human annotations}\label{sec:annotators}
In the absence of ground truth data, we requested five individuals labelers to review the videos in the dataset and provide an aggregate group synchrony score based on their perception of how well the group was interacting. The scores were in the range of 1 to 5, with 1 implying the group was completely unsynchronized, 5 being completely synchronized. The labelers were instructed to judge overall synchrony so that even if two people in the group are interacting well with each other, but not with the third person, the group could not receive a high score. To account for the subjectivity in the labeling, we computed the total variances for the scores provided by all but one of the labelers (we did this five times), and removed the set of annotations that caused the largest variance in the set. We also spot-checked the variances across each group and when this was larger than a preset threshold,  had additional ad hoc labelers re-score the video - this was done for only two groups in the entire dataset. The average score obtained from  the labelers was now considered as the human impression of synchrony, the new gold-standard we used to train the network. 

\subsection{LSTM Model}\label{sec:model}
\begin{figure}[ht]
	\centering
	\includegraphics[width= 0.99\linewidth]{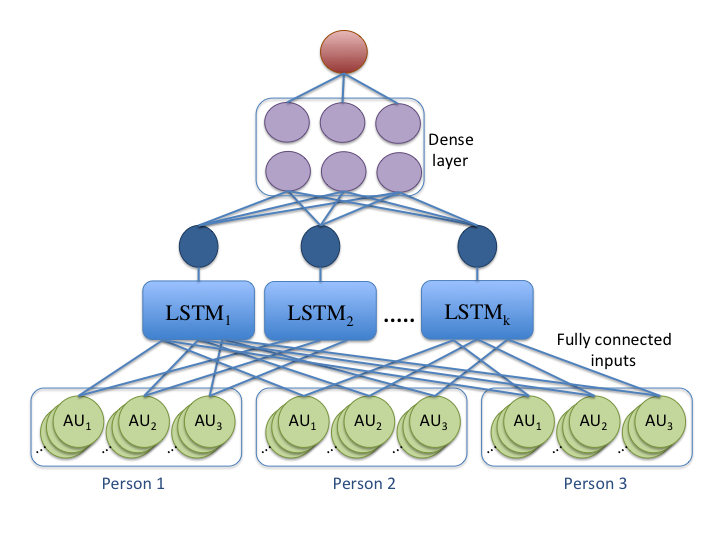}
	\caption{An LSTM network to predict the synchrony among 3 sets of input signals, where each set has 3 channels. For the first layer, \underline{not} all the connections are shown in the diagram for the sake of clarity, but every channel of data from each person is connected to all the input nodes of every LSTM network. Note: the number of LSTM networks does not necessarily equal the number of sets at the input.}
	\label{fig:lstm2}
\end{figure}
We implement an LSTM network followed by a fully connected layer to predict the extent of interaction between two or more sets of input signals. The network utilizes a lookback feature of one second (or 30 frames).  It is important to note that the number of LSTM networks does not necessarily have to be equal to the number of sets at the input. More information is provided on how we selected the optimal number of input networks in Section \ref{sec:testing}. All neurons use the ReLU activation function. For the real-life data, since we are analyzing 3 different action units (the 3 most prominently changing AUs), in the first layer, each LSTM network has a neuron for each channel of the input for all individuals.  A simplified model can be seen in Figure \ref{fig:lstm2}.

\section{Testing and Results}\label{sec:testing}
In this section, we first define the different performance measures used to evaluate the architecture and then describe the overall testing methodology, both for the synthetic and the real datasets. Lastly, we provide the results obtained from both sets of experiments.

\subsection{Performance measures}
To test the effectiveness of our model, we computed the mean of the absolute percent error $\mu_e$ given as: \\
$$\mu_e = \frac{1}{N}\sum_{i=1}^N \left| \frac{Y_i - \hat{Y_i}}{Y_i} \right|$$ \\

Additionally, we computed the standard deviation of percent error $\sigma_e$ given as:  \\
$$\sigma_e = \sqrt{\frac{1}{N}\sum_{i=1}^N \left(\frac{Y_i - \hat{Y_i}}{Y_i} - \mu\right)^2}$$\\

Finally, we computed the standard statistical measure, $R^2$ also known as the coefficient of determination, as:\\
$$R^2 = 1-\frac{\sum_{i=1}^N (Y_i - \hat{Y_i})^2}{\sum_{i=1}^N Y_i - \mu}$$\\

\noindent  where $Y_i$ is the ground-truth synchronicity measure (or measure of cohesion) of the $i^{th}$ group, $\hat{Y_i}$ is the model's prediction on the same $i^{th}$ group and $N$ is the total number of groups. 

For synthetic data, the ``group'' is the pair of signals generated using the \texttt{DATAGEN} algorithm. For the more complex real-life scenario, the group consists of the data from the three people in conversation.  Each person's data consists of 3 AUs.

\subsection{Testing methodology}
The testing methodologies are similar for the synthetic and the real-life datasets, where the time series data is input to an LSTM architecture, the output codes from all the LSTMs are fed into a fully-connected dense neural network and the algorithm is trained using a regression-based mean-squared error (MSE) loss (Figure \ref{fig:lstm2}).

As proof-of-concept, we first tested the network on the synthetic dataset to attempt to recover the covariate measure between the two interacting signals (this is known since we generated the data). We trained the network on 100 pairs of signals, each signal having a length of 1000. Using an overlapping window of length 100 and a stride of 1, we extracted pairs of sub-signals which were fed into the LSTM architecture where 80\% was used for training and 20\% for validation. During training, each pair of sub-signals extracted with the overlapping window is input to the network and labeled with the overall synchrony metric which in this case is the covariate value between the entire signals. 

\begin{figure}[ht]
	\centering
	\includegraphics[width= 0.9\linewidth]{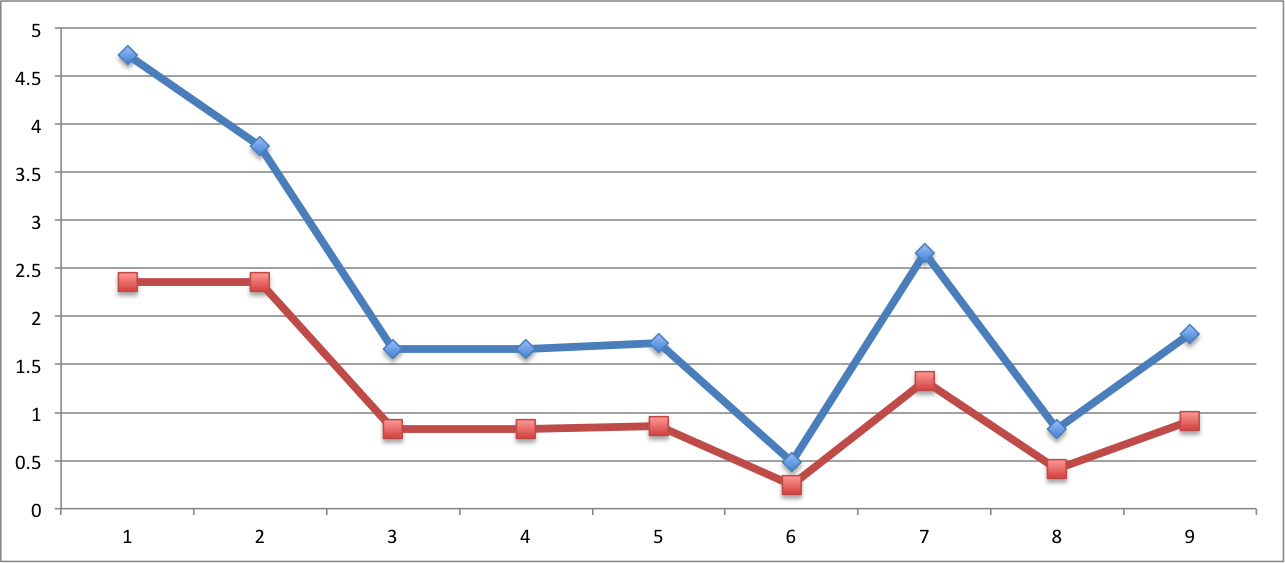}
	\caption{x-axis is the number of LSTM networks and y-axis is the lowest error obtained from the model on the same data. The red (or lower curve) represents the results obtained from training data and the blue (or upper) represents the results obtained from the validation data.}
	\label{fig:lstm_nums}
\end{figure}

Since we selected the number of LSTM networks arbitrarily, we tested using a varying number of LSTM networks, from 1 to 9 as shown in Figure \ref{fig:lstm_nums}. The optimal number of LSTMs for the synthetic dataset was 6, although it can be seen that the number of LSTMs makes only a small difference to the end result.

We tested the network on 100 pairs of signals, each of length of 1000. Similar to the training paradigm, the pairs of sub-signals were extracted and fed to the network to retrieve their synchrony metric. The overall mean error between the predicted and actual values was \textbf{0.5\%}.

For the real-life data, the same paradigm was employed. All signals were one-minute or 1800 frames long. Similar to the process employed for the synthetic data, we used an overlapping window of size 30 and fed the sub-signals to a similar LSTM architecture. In this case, we used 10 LSTM networks and the inputs consisted of 3 sets of 3 channels each. During training, the sub-signals were labeled with the ground-truth label (previously assigned by human annotators to the entire group of three, as a measure of their perceived level of interaction - between 1 and 5). 

To train, we used 5-fold cross validation across the groups \ie 24 groups were used for training and the remaining 8 for testing, over 5 folds. As a type of control baseline, to ensure we are accurately measuring interactional synchrony and not some other underlying phenomenon, for each group in the test set, we selected one of the three original members and combined their data with that of two individuals from two other randomly selected groups to create a new group (referred to as \emph{random} in Table \label{tab:res} since we should be unable to predict better than random). We then performed the predictions on these permutations and compared the predicted values to the ground-truth labels of the group. After training the regressor, we obtained results within a mean of $2.96\%$ error when performing 5-fold validation with a $30^{th}$ order Markov assumption as compared to 26.1\% on the random permutations.  The full set of results on the real-life dataset are shown in Table \label{tab:res}.

\begin{table}[h]
	\centering
	\begin{tabular}{|c|c|c|c|}
		\hline
		\rowcolor{lightgray}  &  & Mean of Absolute & Standard Deviation \\
		\rowcolor{lightgray} Data & $R^2$ & Percent Error & Percent Error\\
		\hline      
		Random & $0.00717$ & $0.261$ & $0.203$\\
		5-Fold validation  &$0.979$ & $0.0296$ & $0.0296$\\
		\hline
	\end{tabular}
	\caption{Results after running 5 fold validation on dataset and on control baseline}
	\label{tab:res}
\end{table}

The $R^2$ value is the most telling metric in this work, where a value of $1.0$ indicates that the model fits the data well and the differences between the ground-truth values and the model's predicted values are small and unbiased. An $R^2$ value of $0.0$ indicates the contrary. For this work, on the real-life data set, after five folds of validation, the resulting $R^2$ was \textbf{0.979} while the value for the random permutation was \textbf{0.00717}.

\section{Conclusion}
We can observe that our model is highly accurate. Furthermore, based on the analysis and results reported, we can be confident the proposed architecture is looking to measure synchrony. Unfortunately, as with many techniques based on neural networking, it is difficult to completely understand the inner workings of this model and fully interpret what the architecture is inferring from the data. Thus, we can only conjecture about the means to which it derives its conclusions.
Notwithstanding, this type of architecture can be used for efficient classification of group dynamics. Potential applications using transfer learning and applying this to sequences could be investigated.

\bibliographystyle{ieee}
\bibliography{bmvc_review}
\end{document}